\documentclass{Interspeech}

\interspeechcameraready

\title{Face2VoiceSync: Lightweight Face-Voice Consistency for \\ Text-Driven Talking Face Generation 
}

\author[affiliation={1}]{Fang}{Kang}
\author[affiliation={2}]{Yin}{Cao}
\author[affiliation={1}]{Haoyu}{Chen}

\affiliation{Center for Machine Vision and Signal Analysis}{University of Oulu}{Finland}
\affiliation{Xi’an Jiaotong Liverpool University}{China}

\email{fang.s.kang@gmail.com, yin.k.cao@gmail.com, chen.haoyu@oulu.fi}


\usepackage{comment}
\usepackage[usenames, dvipsnames]{color}
\usepackage{amsmath,amsfonts}
\usepackage[caption=false,font=normalsize,labelfont=sf,textfont=sf]{subfig}
\usepackage{textcomp}
\usepackage{hyperref}
\usepackage{stfloats}
\usepackage{verbatim}
\usepackage{array}
\usepackage{enumitem}
\usepackage{soul}
\usepackage{cite}
\usepackage{siunitx}
\usepackage{adjustbox}
\usepackage{hhline}
\usepackage{makecell}
\usepackage{caption}
\usepackage{float}
\usepackage{steinmetz}
\usepackage{tablefootnote}
\usepackage{scrextend}
\usepackage{multirow}
\usepackage{multicol}
\usepackage{url, times, booktabs, tabularx, amssymb, bm, bbm}
\usepackage{algorithm, algorithmic}
\usepackage{bbding}
\usepackage{graphicx}
\usepackage{ragged2e}
\usepackage{tcolorbox}

\begin{document}

\maketitle

\begin{abstract}

Recent studies in speech-driven talking face generation achieve promising results, but their reliance on fixed-driven speech limits further applications (e.g., face-voice mismatch). Thus, we extend the task to a more challenging setting: given a face image and text to speak, generating both talking face animation and its corresponding speeches. Accordingly, we propose a novel framework, Face2VoiceSync, with several novel contributions: 1) Voice-Face Alignment, ensuring generated voices match facial appearance; 2) Diversity \& Manipulation, enabling generated voice control over paralinguistic features space; 3) Efficient Training, using a lightweight VAE to bridge visual and audio large-pretrained models, with significantly fewer trainable parameters than existing methods; 4) New Evaluation Metric, fairly assessing the diversity and identity consistency. Experiments show Face2VoiceSync achieves both visual and audio state-of-the-art performances on a single 40GB GPU.

\end{abstract}

\section{Introduction}


Talking Face Generation (TFG) aims to synthesize realistic facial animations synchronized with speech, enabling applications in virtual avatars, digital assistants, and media production. Recent advancements, particularly diffusion-based methods, have significantly improved lip-sync accuracy and facial expressiveness, drawing increasing research interest.



However, most existing approaches are based on audio-driven pipelines \cite{wang2021audio2head, zhang2023sadtalker, ma2023dreamtalk, xu2024hallo,tan2024edtalk}, which generate expressive facial movements and accurate lip synchronization from \textbf{pre-recorded fixed audio}, limiting their flexibility and leading to potential mismatches between voice and facial identity.  To address this, we explore a more challenging setting: \textit{generating both talking faces and corresponding speech from a single static image and textual input}, ensuring voice-identity consistency and enhanced controllability.

To our knowledge, the closest setting among existing methods to our proposed task is \textit{text-driven TFG} that generates speech directly from text, enabling precise control over speech content, and offering greater flexibility \cite{zhang2022text2video, jang2024faces}. Existing methods can be divided into two categories. The first employs a cascaded pipeline \cite{zhang2022text2video, wang2023text, ye2023ada}, where TFG models are followed by text-to-speech (TTS) systems. The second approach involves latent space fusion \cite{jang2024faces,Mitsui2023,choi2024text}, where speech and video are generated simultaneously within a shared latent space. Beyond improving speech controllability, text-driven methods can also enhance face-voice consistency by incorporating facial information into speech generation, ensuring that the synthesized voice aligns with the speaker’s visual appearance. This provides an advantage over audio-driven approaches, which may struggle to fully align visual and auditory features. On the other hand, current methods \cite{jang2024faces,wang2023text, liu2022generating} frequently conceptualize the face-voice relationship as a one-to-one mapping, aiming to establish a direct correlation between facial features and voice attributes to improve the naturalness of generated speech.


While text-driven methods offer considerable promise and benefits, they also pose several challenges in real-world implementations. First, generating high-quality audio and video simultaneously requires significant computational resources when training large-scale multimodal models from scratch \cite{ye2023ada}. Although adopting pre-trained unimodal models can reduce this burden, maintaining cross-modal alignment remains a challenge. Second, generating corresponding voice characteristics based on facial appearance is barely considered. The existing methods by default assume that voice generation is restricted to a single fixed voice style per face, limiting models' adaptability to unseen faces (i.e., it's desirable that models can generate diverse voices with a given face for uses to choose from). Third, as a result of the above assumption (fixed voice style per face), existing methods \cite{choi2024text, diao2023ft2tf} primarily focus on video quality rather than face-voice identity consistency, with few efforts have been made for standardized metrics to evaluate this consistency.



\begin{figure}[t]
    \centering
    \includegraphics[width=1.0\linewidth]{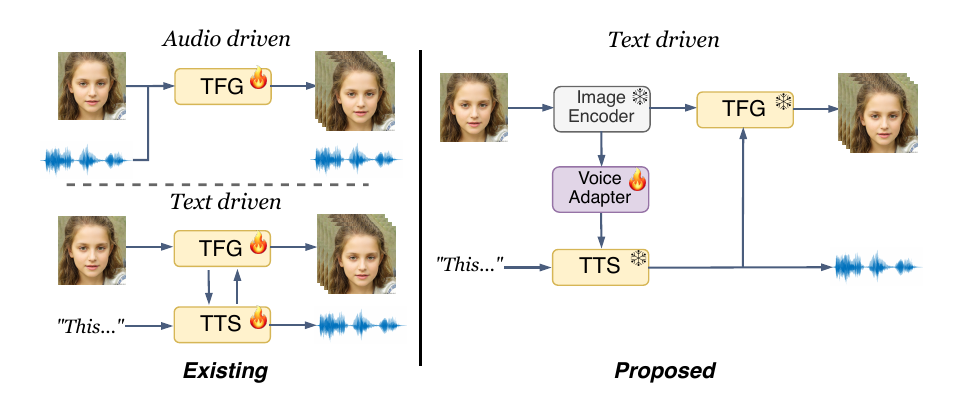}
    \caption{A framework comparison between Face2VoiceSync and other existing methods.}
    \label{fig:comparisonframework}
    \vspace{-0.5cm}
\end{figure}

Towards the challenges discussed above, we propose Face2VoiceSync, a scalable method for the TFG task by improving cross-modal alignment with flexible voice customization. An overview of the Face2VoiceSync framework is shown in Figure \ref{fig:comparisonframework}. Specifically, we make several unique contributions: 1) Voice-Face Alignment, we use a VoiceAdapter module following the face image encoder to ensure generated
voices match facial appearance, 2) we adopt a lightweight VAE, enabling diverse face-based voice generation rather than restricting voice generation to a single fixed voice style per face. 3) Instead of modifying the frozen video generator, we fine-tune the latent space for better cross-modal consistency and zero-shot adaptation. 4) Additionally, we introduce the Diversity-Consistency Tradeoff Score (DCTS) to fairly evaluate the generated voice diversity and identity consistency. By leveraging pre-trained unimodal models and VAE-driven alignment, Face2VoiceSync achieves state-of-the-art performance with reduced training costs. Demos are available\footnote{https://face2voicesync.github.io/}.



\section{Related works}

\textbf{Talking face generation (TFG)} has predominantly utilized audio-driven methods to create speaking portraits from a static image and audio input. Early works, such as Chen et al.  \cite{chen2019hierarchical} and Prajwal et al. \cite{prajwal2020lip} convert speech into mouth movements, laying the foundation for TFG. SadTalker \cite{zhang2023sadtalker} improved realism using implicit 3D coefficient modulation, while diffusion-based models like GAIA \cite{he2024gaia} and DreamTalk \cite{ma2023dreamtalk} enhanced lip synchronization and speaking style diversity. More recently, Hallo \cite{xu2024hallo} introduced hierarchical cross-attention to better synchronize audio with non-identity-related facial motions, enabling more natural and expressive talking faces.

Beyond audio-driven methods, text-driven talking face generation offers greater flexibility by eliminating reliance on audio input, generating both speech and facial animations directly from text. A common approach integrates TTS and TFG models, though methods such as ADA-TTA \cite{ye2023ada} still require reference audio. To improve cross-modal synchronization, Zhang et al. \cite{zhang2022text2video} and Wang et al. \cite{wang2023text} introduce intermediate speech representations, while Jang et al. \cite{jang2024faces} and Mitsui et al. \cite{Mitsui2023} propose direct text-to-animation mappings. However, the absence of audio input poses challenges for natural synchronization, as models lack essential prosodic and temporal cues \cite{nazarieh2024portraittalk}.

\noindent \textbf{Conditional text-to-speech (TTS) synthesis} generates speech from text while conditioning on various inputs, including text, audio, and images. CosyVoice \cite{du2024cosyvoice} enables zero-shot voice cloning by extracting speaker characteristics from reference audio but still requires target speaker samples. For TFG, image-conditioned methods provide a more suitable alternative by modeling speaker traits directly from facial images. Lee et al. \cite{lee2023imaginary} train a TTS model using facial features but face challenges in duration prediction and speech quality. Lee et al. \cite{lee2024fvtts} further enhance personalized voice synthesis with a face encoder, achieving better gender differentiation. However, their approach still struggles to capture fine-grained vocal traits and lacks precise modulation of paralinguistic attributes in generated speech.

In this work, we propose Face2VoiceSync, which improves cross-modal alignment and enables flexible voice customization using a lightweight VAE. By mapping facial features and speech into a shared latent space, it captures fine-grained vocal traits and refining paralinguistic modulation.

\section{Method}  
\begin{figure}[t]
\centering
\includegraphics[width=1.0\linewidth]{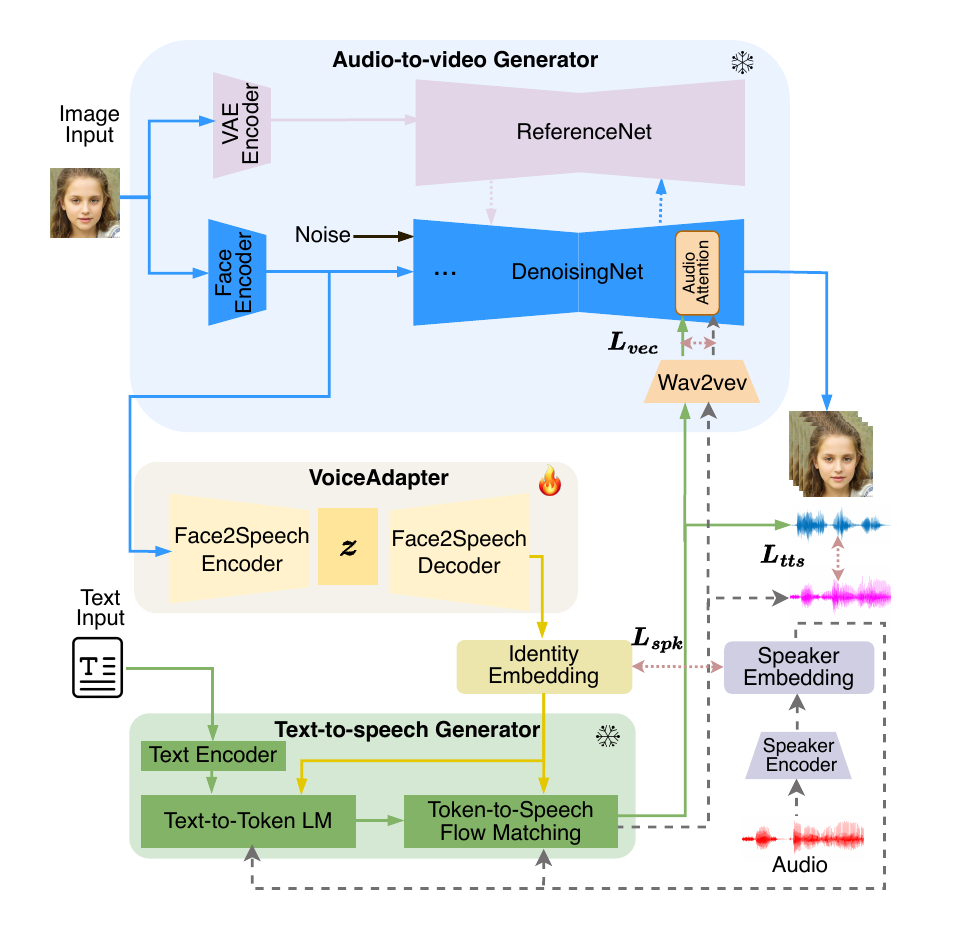}
\caption{Framework of the proposed text-driven talking face generation system. The framework takes both text and image inputs, leveraging VoiceAdapter to bridge the audio-to-video generator and text-to-speech generator. Dashed lines represent data flow paths that are only active during the training phase.}
\label{fig:framework}
\vspace{-0.5cm}
\end{figure}



We propose Face2VoiceSync, a scalable text-driven talking face generation method. The overall pipeline is illustrated in Figure \ref{fig:framework}. In Section 3.1, we provide an overview of the Face2VoiceSync framework. Section 3.2 details VoiceAdapter and its training strategies. Finally, Section 3.3 introduces DCTS, our proposed evaluation metric.

\subsection{Face2VoiceSync Framework}

Face2VoiceSync generates natural and realistic talking face videos with editable speech content and voice characteristics. It employs a lightweight pre-training strategy, leveraging frozen pre-trained video generators and large-scale TTS models. In the following, we introduce each components in the framework based on the data flow (a given image and the text to speak).

\noindent \textbf{Text-to-Speech Generation}.
With the given text, to enable flexible speech editing, we use the pretrained model CosyVoice \cite{du2024cosyvoice}, a large-scale LLM-based TTS model. CosyVoice consists of an LLM-based text-to-token generator and a conditional flow matching model for token-to-speech synthesis, ensuring high-quality, natural, and zero-shot speech generation. The generated speech then serves as the driving signal for the video generator.

\noindent \textbf{Audio-to-Video Generation}. 
Given a face image, we adopt Hallo \cite{xu2024hallo}, a frozen diffusion-based model known for producing expressive and realistic facial animations. As shown in Figure \ref{fig:framework}, the input image is processed through ReferenceNet and DenoisingNet, where ReferenceNet guides DenoisingNet throughout the generation process. The driving audio interacts with non-identity-related facial motions (e.g., lip movements, expressions, and poses) via cross-attention, ensuring precise synchronization between speech and facial animation.

\noindent \textbf{Cross-Modal Alignment via VoiceAdapter}. Directly fusing the video generator and CosyVoice, which are trained separately, leads to cross-modal mismatches, while joint training incurs high computational costs. To address this, we introduce VoiceAdapter (detailed illustration as below), a lightweight VAE that maps facial features to consistent identity embeddings. Instead of a fixed one-to-one mapping, it learns a distribution-to-distribution transformation through sampling, enabling many-to-many mapping. This enhances zero-shot adaptability and ensures coherent vision-audio synchronization.


\subsection{VoiceAdapter}
VoiceAdapter comprises an encoder, a sampling module, and a decoder, with both the encoder and decoder consisting of two linear layers. It models the transformation as a conditional probability distribution \( p(s | v) \), where \( v \) is the facial feature embedding, and \( s \) is the corresponding generated identity embedding. To ensure compatibility with pre-trained modules, we use latent features from the video generator’s face encoder as input to the VAE encoder, mapping them to a latent distribution as \( q(z | v) \), and sample \( z \sim q(z | v) \). Within this space, the VAE enables face-to-voice mapping as sampling from a distribution conditioned with facial feature embeddings rather than a fixed mapping from a point. This means that the model is able to sample diverse identity representations (generate different voice styles) over a coherent distribution instead of point sampling (fixed voice style), improving generalization to unseen faces and supporting zero-shot adaptability.

To regularize the latent space, we apply Maximum Mean Discrepancy (MMD) regularization \cite{gretton2012kernel} to ensure that the posterior distribution remains close to a Gaussian prior:
\begin{equation}
D_{\text{MMD}}(q(z|v), p(z)) \rightarrow 0.
\end{equation}
This helps maintain a balance between voice diversity and identity consistency. The VAE decoder then regenerates \( s \) from \( z \) with a conditional probability \( p(s | z) \), producing identity embeddings that serve as conditional inputs to CosyVoice for speech generation.

VoiceAdapter optimization consists of two stages: the embedding learning stage and the generative pre-training stage. In the embedding learning stage, it learns identity embeddings by extracting speaker features from facial features in the latent space. Facial features are obtained from a frozen face encoder in TFG, while target speaker embeddings come from the TTS speaker encoder \cite{wang2023cam++}. The optimization follows the loss function:

\begin{equation}
\mathcal{L}_{\text{Spk}} =  \lambda_{\text{Rec}} \mathcal{L}_{\text{Rec}} + \lambda_{\text{Con}} \mathcal{L}_{\text{Con}} + \lambda_{\text{cen}} \mathcal{L}_{\text{cen}},
\end{equation}
where \( \lambda_{\text{Rec}}, \lambda_{\text{Con}}, \lambda_{\text{cen}} \) are weights balancing the loss terms. The reconstruction loss \( \mathcal{L}_{\text{Rec}} \) is implemented using cosine similarity to align identity embeddings with the target speaker embedding. The contrastive loss \( \mathcal{L}_{\text{Con}} \) enhances identity discrimination. The dynamic center loss \( \mathcal{L}_{\text{cen}} \) stabilizes identity embeddings by reducing intra-class variance and preserving inter-class distinction, defined as:
\begin{equation}
\mathcal{L}_{\text{cen}} = \frac{1}{2} \sum_{n=1}^N  \Vert {s}_n - c_{y_n}  \Vert^2,
\end{equation}
where \( N \) is the batch size, \( {s}_n \) is the predicted identity embedding of sample \( n \), and \( c_{y_n} \) is the corresponding class center, obtained as follows:

\begin{equation}
c_{y_n} \leftarrow \alpha \cdot {s}_n + (1 - \alpha) \cdot c_{y_n},
\end{equation}
where \( \alpha \) controls the update rate. 

In the generative pretraining stage, as shown in Figure \ref{fig:framework}, VoiceAdapter connects to the frozen LLM-based CosyVoice to preserve its generative capabilities, and subsequently to the Wav2Vec encoder \cite{baevski2020wav2vec} to mitigate cross-modal mismatches. The optimization is guided by:
\begin{equation}
\mathcal{L}_{\text{gen}} =  \lambda_{\text{tts}} \mathcal{L}_{\text{tts}} + \lambda_{\text{vec}} \mathcal{L}_{\text{vec}},
\end{equation}
where \( \lambda_{\text{tts}}  \) and \( \lambda_{\text{vec}}  \) control the contributions of each loss term. \( \mathcal{L}_{\text{tts}} \) is the cosine similarity loss between the speaker embeddings of the image-conditioned and target-audio-conditioned CosyVoice outputs, ensuring speaker consistency. \( \mathcal{L}_{\text{vec}} \) is an MSE loss between Wav2Vec representations, reinforcing audiovisual synchronization in audio-to-video generator's cross-attention mechanism.


After two-stage training, VoiceAdapter effectively learns identity-related features while ensuring compatibility across the pre-trained TTS and video generation models. This alignment enables identity-consistent speech synthesis and seamless integration within the Face2VoiceSync framework.

\subsection{Diversity-Consistency Tradeoff Score (DCTS)}  

To provide a more comprehensive evaluation of face-to-speech synthesis, we carefully design DCTS, a novel metric that quantifies the relative disparity between intra-class and inter-class distributions. By considering both probabilistic and geometric perspectives, DCTS offers a holistic measure of identity consistency and diversity in generated speech, ensuring a robust assessment.

From a probabilistic perspective, human voices are shaped by physiological structures, making different speakers inherently independent while maintaining strong intra-speaker dependencies—a key assumption in blind speech separation \cite{kang2019low}. Under this assumption, we assess face-to-voice mapping by quantifying the difference between intra-class and inter-class independence using the Relative Independence Ratio (RIR), defined as:

\begin{equation}
\text{RIR}_m = \frac{I_{m,\text{intra}}}{I_{m,\text{inter}} + \epsilon},
\end{equation}
where  \( I_{m,\text{intra}} \) and \( I_{m,\text{inter}} \) denote the independence of intra-class and inter-class samples, respectively. These values are computed using the Kullback-Leibler (KL) divergence between joint and marginal distributions:
\begin{equation}
    I_m = \text{KL}\left(p(s_{1}, s_{2}, \dots, s_{K}) \Vert \prod_{k=1}^{K} q(s_{k})\right),
\end{equation}
where \( K \) is the number of samples, \( q \) represents the marginal probability density of each sample, and \( p \) denotes the joint probability distribution among  \( K \)  samples.

From a geometric perspective, similar speech identity samples should have smaller feature distances, while dissimilar ones should be farther apart. To quantify this, we introduce the Relative Cosine Ratio (RCR), which measures the difference between intra-class and inter-class distances:

\begin{equation}
\text{RCR}_{m} = \frac{D_{m,\text{inter}}}{D_{m,\text{intra}} + \epsilon}.
\end{equation}
where  \( D_{m,\text{intra}} \) and \( D_{m,\text{inter}} \) denote the cosine distances of intra-class and inter-class samples, respectively. 

To integrate both probabilistic and geometric perspectives into a unified evaluation, we normalize and combine RIR and RCR into a single metric:
\begin{equation}
\text{DCTS} = \sqrt{\frac{\text{RIR}_m}{\text{RIR}_m + 1} \cdot \frac{\text{RCR}_m}{\text{RCR}_m + 1}}.
\end{equation}

This formulation captures both statistical independence and geometric distance properties, providing a robust measure of identity consistency and diversity in face-to-voice mapping. When intra-class and inter-class variance are indistinguishable, DCTS = 0.5. A DCTS \( >\) 0.5 indicates well-clustered intra-class representations, ensuring accurate alignment with the target identity, while DCTS \( < \) 0.5 reflects excessive intra-class dispersion, suggesting a deviation from the target. For a well-aligned mapping, DCTS should be greater than 0.5. A higher DCTS implies stronger identity consistency but lower diversity, whereas a lower DCTS indicates greater diversity at the cost of consistency.

\section{Experiments} 
\subsection{Experimental Settings}

\begin{table}[t]
\centering
\caption{TFG performance comparisons with existing talking face generation approaches on the HDTF dataset. Lower is better for FID, FVD, Sync-D, while higher is better for Sync-C.}
\label{tab:quantitative_comparison}
\resizebox{\linewidth}{!}{%
\begin{tabular}{lcccc}
\toprule
Method       & FID $\downarrow$  & FVD $\downarrow$  & Sync-C $\uparrow$  & Sync-D $\downarrow$  \\
\midrule
Hallo \cite{xu2024hallo}       & \textbf{20.545}           &  \textbf{173.497}       &   7.750            &   7.659                  \\
SadTalker \cite{zhang2023sadtalker}     & 22.340            &  203.860          & \textbf{7.885 }             & \textbf{7.545}                       \\
Ours         & 24.272           &  194.462          & 7.605              & 7.876                      \\
Real video   &  \( -\)                  &  \( -\)                  & 8.700              & 6.597                           \\
\bottomrule
\end{tabular}%
}
\end{table}


We conduct our experiments using LRS2 \cite{son2017lip} and HDTF \cite{zhang2021flow} datasets. The LRS2 dataset contains talking face videos paired with transcription labels. We split it into 3,347 speakers for training and 318 speakers for testing. HDTF consists of high-quality YouTube videos with more than 300 speakers, which we use for testing. For a fair comparison, all the compared methods are tested on the same audio samples for audio-driven testing and text sentences for TTS evaluation.


The encoder and decoder of VoiceAdapter each consist of two linear hidden layers, ensuring a lightweight architecture. The model takes 512-dimensional face features, extracted from the face encoder \cite{xu2024hallo}, as input and generates 192-dimensional identity embeddings. During training, three frames are randomly sampled from each video as image inputs. DCTS computation is applied to the speaker embeddings generated by the speaker encoder \cite{wang2023cam++}, with probability densities estimated using Kernel Density Estimation (KDE) \cite{parzen1962estimation}.  The model is optimized with Adam \cite{kingma2014adam} at a learning rate of \(5 \times 10^{-5}\), and training is performed using PyTorch on one single 40GB A100 GPU.

For TFG, we conducted a comparative analysis of our proposed method against publicly available implementations of Hallo \cite{xu2024hallo} and SadTalker \cite{zhang2023sadtalker}. For TTS, we conducted a comparative analysis of our proposed method against Cosyvoice and FaceTTS \cite{lee2023imaginary}. 

\subsection{Experiment Results}
\textbf{Talking Face Generation.} We compare our framework with Hallo \cite{xu2024hallo} and SadTalker \cite{zhang2023sadtalker} on the HDTF dataset using Fréchet Inception Distance (FID), Fréchet Video Distance (FVD), Synchronization-C (Sync-C), and Synchronization-D (Sync-D), as shown in Table~\ref{tab:quantitative_comparison}. Our method achieves performance comparable to audio-driven approaches, demonstrating its effectiveness in mitigating cross-modal mismatches. This enables high-quality talking face generation using only a single image and editable text. The improved alignment is partly due to the alignment of Wav2Vec outputs used in the TFG module, further reducing modality mismatches.

\noindent \textbf{Conditional TTS.}
We compare our method with CosyVoice \cite{du2024cosyvoice} and FaceTTS \cite{lee2023imaginary} on the LRS2 dataset using: Word Error Rate (WER) and Speaker Similarity (SS) from \cite{du2024cosyvoice} (ASR-based and speaker embedding similarity), and our proposed DCTS metric with sub-metrics \(\text{RIR}_m\) and \(\text{RCR}_m \). The results are presented in Table~\ref{tab:audio_performance_comparison}. Our method achieves the lowest WER, demonstrating superior speech intelligibility. CosyVoice, however, shows a higher error rate than its reported (\(\sim \) 3\%)  performance, likely due to the short reference audio in LRS2, highlighting its reliance on sufficient input—a limitation our image-conditioned approach avoids. In terms of DCTS, CosyVoice scores the highest, reflecting strong intra-speaker consistency. FaceTTS, while achieving a reasonable SS score, has a low DCTS close to 0.5, suggesting reduced speaker distinction. This indicates that its generated voices have limited differentiating ability across identities. Our method balances consistency and diversity, maintaining strong face-voice consistency while offering greater variation than CosyVoice. The aligned trends in  \(\text{RIR}_m\)and \(\text{RCR}_m \) further validate their contribution to DCTS, demonstrating its evaluating robustness.

\begin{table}[t]
\centering

\caption{TTS performance comparison with CosyVoice and FaceTTS on the LRS2 dataset.
Lower is better for WER, while higher SS, DCTS, \(\text{RIR}_m\) and \(\text{RCR}_m\) reflect more consistent, lower ones reflect more diverse.}
\label{tab:audio_performance_comparison}
\resizebox{\linewidth}{!}{%
\begin{tabular}{lccccc}
\toprule
Method & WER(\%)$\downarrow$ & SS $\uparrow$ & DCTS  & \(\text{RIR}_m\) & \(\text{RCR}_m\)   \\
\midrule
CosyVoice \cite{du2024cosyvoice} & 10.739  & \textbf{0.712} & 0.747 & 1.684 & 7.084 \\
FaceTTS \cite{lee2023imaginary}  & 19.809  & 0.643 & 0.508 & 0.8193 & 1.358 \\
Ours                          & \textbf{3.669} &{0.669} & 0.659 & {1.231} & {3.554} \\
\bottomrule
\end{tabular}%
}
\end{table}



\subsection{Ablation Studies}

\begin{table}[t]
\centering
\caption{ Quantitative ablation results.}
\label{tab:Ablation}
\resizebox{\linewidth}{!}{%
\begin{tabular}{lccccc}
\toprule
Method & WER(\%)$\downarrow$ & SS $\uparrow$ & DCTS  & \(\text{RIR}_m\) & \(\text{RCR}_m\)   \\
\midrule
Ours                                   & \textbf{3.669} &{0.669} & 0.659 & {1.231} & {3.554}  \\
w/o VAE                                &  4.423  & \textbf{0.672} & 0.662  & 1.328 & 3.316 \\
w/o  \( \mathcal{L}_{\text{Gen}} \)    & {3.842} &{0.653} & 0.6449 &  1.167 & 3.392 \\

GMM-based estimation & \( -\)  & \( -\) & 0.664 & {1.298} & {3.554} \\

\bottomrule
\end{tabular}%
}
\end{table}


We conduct an ablation study, summarized in Table 3, to assess the impact of TTS quality on video generation.


First, we compare our method to a baseline that uses a two-layer MLP to map facial features to speaker embeddings. Removing the VAE increases WER (3.669 → 4.423), indicating that the VAE improves speech intelligibility by learning better identity representations. The MLP's fixed mapping restricts diversity while promoting tighter intra-class clustering.


Next, we compare models with and without \( \mathcal{L}_{\text{Gen}} \). Excluding this loss slightly increases WER and decreases DCTS (0.659 → 0.6449), suggesting that \( \mathcal{L}_{\text{Gen}} \) helps align identity embeddings with the TTS model while maintaining consistency.


Finally, we compare KDE and GMM-based estimation for DCTS computation. The results show that changes in \(\text{RIR}_m\)  are minimal, as we use relative values. Combining  \(\text{RIR}_m\)  with  \(\text{RCR}_m\) further reduces differences, minimizing their effect on DCTS and demonstrating the robustness of our metric.

\section{Conclusion}

We propose Face2VoiceSync, a scalable text-driven talking face generation method that synthesizes natural, identity-consistent animation and speech from a single image and text. Unlike prior works assuming a fixed face-to-voice mapping, we model it as a probability distribution problem, capturing natural voice variability. A lightweight VAE bridges the modality gap, refining facial features into identity embeddings for better cross-modal alignment with strong zero-shot adaptability and voice diversity while reducing training costs. Additionally, we introduce DCTS, a novel metric for evaluating voice diversity and consistency, extendable to other many-to-many mapping tasks. Experiments confirm Face2VoiceSync achieves state-of-the-art performance in both visual and audio aspects.

\section{Acknowledgements}
This work was supported by the University of Oulu and the Research Council of Finland (PROFI7 grant 352788). The authors also acknowledge CSC-IT Center for Science, Finland, for providing computational resources.


\bibliographystyle{IEEEtran}
\bibliography{mybib}

\begin{thebibliography}{10}
\providecommand{\url}[1]{#1}
\csname url@samestyle\endcsname
\providecommand{\newblock}{\relax}
\providecommand{\bibinfo}[2]{#2}
\providecommand{\BIBentrySTDinterwordspacing}{\spaceskip=0pt\relax}
\providecommand{\BIBentryALTinterwordstretchfactor}{4}
\providecommand{\BIBentryALTinterwordspacing}{\spaceskip=\fontdimen2\font plus
\BIBentryALTinterwordstretchfactor\fontdimen3\font minus \fontdimen4\font\relax}
\providecommand{\BIBforeignlanguage}[2]{{%
\expandafter\ifx\csname l@#1\endcsname\relax
\typeout{** WARNING: IEEEtran.bst: No hyphenation pattern has been}%
\typeout{** loaded for the language `#1'. Using the pattern for}%
\typeout{** the default language instead.}%
\else
\language=\csname l@#1\endcsname
\fi
#2}}
\providecommand{\BIBdecl}{\relax}
\BIBdecl

\bibitem{wang2021audio2head}
S.~Wang, L.~Li, Y.~Ding, C.~Fan, and X.~Yu, ``Audio2head: Audio-driven one-shot talking-head generation with natural head motion,'' \emph{arXiv preprint arXiv:2107.09293}, 2021.

\bibitem{zhang2023sadtalker}
W.~Zhang, X.~Cun, X.~Wang, Y.~Zhang, X.~Shen, Y.~Guo, Y.~Shan, and F.~Wang, ``Sadtalker: Learning realistic 3d motion coefficients for stylized audio-driven single image talking face animation,'' in \emph{Proceedings of the IEEE/CVF Conference on Computer Vision and Pattern Recognition}, 2023, pp. 8652--8661.

\bibitem{ma2023dreamtalk}
Y.~Ma, S.~Zhang, J.~Wang, X.~Wang, Y.~Zhang, and Z.~Deng, ``Dreamtalk: When expressive talking head generation meets diffusion probabilistic models,'' \emph{arXiv preprint arXiv:2312.09767}, 2023.

\bibitem{xu2024hallo}
M.~Xu, H.~Li, Q.~Su, H.~Shang, L.~Zhang, C.~Liu, J.~Wang, Y.~Yao, and S.~Zhu, ``Hallo: Hierarchical audio-driven visual synthesis for portrait image animation,'' \emph{arXiv preprint arXiv:2406.08801}, 2024.

\bibitem{tan2024edtalk}
S.~Tan, B.~Ji, M.~Bi, and Y.~Pan, ``Edtalk: Efficient disentanglement for emotional talking head synthesis,'' in \emph{European Conference on Computer Vision}.\hskip 1em plus 0.5em minus 0.4em\relax Springer, 2024, pp. 398--416.

\bibitem{zhang2022text2video}
S.~Zhang, J.~Yuan, M.~Liao, and L.~Zhang, ``Text2video: Text-driven talking-head video synthesis with personalized phoneme-pose dictionary,'' in \emph{ICASSP 2022-2022 IEEE International Conference on Acoustics, Speech and Signal Processing (ICASSP)}.\hskip 1em plus 0.5em minus 0.4em\relax IEEE, 2022, pp. 2659--2663.

\bibitem{jang2024faces}
Y.~Jang, J.-H. Kim, J.~Ahn, D.~Kwak, H.-S. Yang, Y.-C. Ju, I.-H. Kim, B.-Y. Kim, and J.~S. Chung, ``Faces that speak: Jointly synthesising talking face and speech from text,'' in \emph{Proceedings of the IEEE/CVF Conference on Computer Vision and Pattern Recognition}, 2024, pp. 8818--8828.

\bibitem{wang2023text}
Z.~Wang, M.~Dai, and K.~Lundgaard, ``Text-to-video: a two-stage framework for zero-shot identity-agnostic talking-head generation,'' \emph{arXiv preprint arXiv:2308.06457}, 2023.

\bibitem{ye2023ada}
Z.~Ye, Z.~Jiang, Y.~Ren, J.~Liu, C.~Zhang, X.~Yin, Z.~Ma, and Z.~Zhao, ``Ada-tta: Towards adaptive high-quality text-to-talking avatar synthesis,'' \emph{arXiv preprint arXiv:2306.03504}, 2023.

\bibitem{Mitsui2023}
K.~Mitsui, Y.~Hono, and K.~Sawada, ``Uniflg: Unified facial landmark generator from text or speech,'' in \emph{Proceedings of the 24th Annual Conference of the International Speech Communication Association (INTERSPEECH)}, 2023, pp. 5501--5505.

\bibitem{choi2024text}
J.~Choi, M.~Kim, S.~J. Park, and Y.~M. Ro, ``Text-driven talking face synthesis by reprogramming audio-driven models,'' in \emph{ICASSP 2024-2024 IEEE International Conference on Acoustics, Speech and Signal Processing (ICASSP)}.\hskip 1em plus 0.5em minus 0.4em\relax IEEE, 2024, pp. 8065--8069.

\bibitem{liu2022generating}
S.~Liu and J.~Hao, ``Generating talking face with controllable eye movements by disentangled blinking feature,'' \emph{IEEE Transactions on Visualization and Computer Graphics}, vol.~29, no.~12, pp. 5050--5061, 2022.

\bibitem{diao2023ft2tf}
X.~Diao, M.~Cheng, W.~Barrios, and S.~Jin, ``Ft2tf: First-person statement text-to-talking face generation,'' \emph{arXiv preprint arXiv:2312.05430}, 2023.

\bibitem{chen2019hierarchical}
L.~Chen, R.~K. Maddox, Z.~Duan, and C.~Xu, ``Hierarchical cross-modal talking face generation with dynamic pixel-wise loss,'' in \emph{Proceedings of the IEEE/CVF conference on computer vision and pattern recognition}, 2019, pp. 7832--7841.

\bibitem{prajwal2020lip}
K.~Prajwal, R.~Mukhopadhyay, V.~P. Namboodiri, and C.~Jawahar, ``A lip sync expert is all you need for speech to lip generation in the wild,'' in \emph{Proceedings of the 28th ACM international conference on multimedia}, 2020, pp. 484--492.

\bibitem{he2024gaia}
T.~He, J.~Guo, R.~Yu, Y.~Wang, J.~Zhu, K.~An, L.~Li, X.~Tan, C.~Wang, H.~Hu \emph{et~al.}, ``{GAIA}: Zero-shot talking avatar generation,'' in \emph{International Conference on Learning Representations (ICLR)}.

\bibitem{nazarieh2024portraittalk}
F.~Nazarieh, Z.~Feng, D.~Kanojia, M.~Awais, and J.~Kittler, ``Portraittalk: Towards customizable one-shot audio-to-talking face generation,'' \emph{arXiv preprint arXiv:2412.07754}, 2024.

\bibitem{du2024cosyvoice}
Z.~Du, Y.~Wang, Q.~Chen, X.~Shi, X.~Lv, T.~Zhao, Z.~Gao, Y.~Yang, C.~Gao, H.~Wang \emph{et~al.}, ``Cosyvoice 2: Scalable streaming speech synthesis with large language models,'' \emph{arXiv preprint arXiv:2412.10117}, 2024.

\bibitem{lee2023imaginary}
J.~Lee, J.~S. Chung, and S.-W. Chung, ``Imaginary voice: Face-styled diffusion model for text-to-speech,'' in \emph{ICASSP 2023-2023 IEEE International Conference on Acoustics, Speech and Signal Processing (ICASSP)}.\hskip 1em plus 0.5em minus 0.4em\relax IEEE, 2023, pp. 1--5.

\bibitem{lee2024fvtts}
M.~Lee, E.~Park, and S.~Hong, ``Fvtts: Face based voice synthesis for text-to-speech,'' in \emph{Proceedings of the 24th Annual Conference of the International Speech Communication Association (INTERSPEECH)}, 2024, pp. 4953--4957.

\bibitem{gretton2012kernel}
A.~Gretton, K.~M. Borgwardt, M.~J. Rasch, B.~Sch{\"o}lkopf, and A.~Smola, ``A kernel two-sample test,'' \emph{The Journal of Machine Learning Research}, vol.~13, no.~1, pp. 723--773, 2012.

\bibitem{wang2023cam++}
H.~Wang, S.~Zheng, Y.~Chen, L.~Cheng, and Q.~Chen, ``Cam++: A fast and efficient network for speaker verification using context-aware masking,'' \emph{arXiv preprint arXiv:2303.00332}, 2023.

\bibitem{baevski2020wav2vec}
A.~Baevski, Y.~Zhou, A.~Mohamed, and M.~Auli, ``wav2vec 2.0: A framework for self-supervised learning of speech representations,'' in \emph{Advances in Neural Information Processing Systems (NeurIPS)}, vol.~33, 2020, pp. 12\,449--12\,460.

\bibitem{kang2019low}
F.~Kang, F.~Yang, and J.~Yang, ``A low-complexity permutation alignment method for frequency-domain blind source separation,'' \emph{Speech Communication}, vol. 115, pp. 88--94, 2019.

\bibitem{son2017lip}
J.~Son~Chung, A.~Senior, O.~Vinyals, and A.~Zisserman, ``Lip reading sentences in the wild,'' in \emph{Proceedings of the IEEE conference on computer vision and pattern recognition}, 2017, pp. 6447--6456.

\bibitem{zhang2021flow}
Z.~Zhang, L.~Li, Y.~Ding, and C.~Fan, ``Flow-guided one-shot talking face generation with a high-resolution audio-visual dataset,'' in \emph{Proceedings of the IEEE/CVF Conference on Computer Vision and Pattern Recognition}, 2021, pp. 3661--3670.

\bibitem{parzen1962estimation}
E.~Parzen, ``On estimation of a probability density function and mode,'' \emph{The Annals of Mathematical Statistics}, vol.~33, no.~3, pp. 1065--1076, 1962.

\bibitem{kingma2014adam}
D.~P. Kingma, ``Adam: A method for stochastic optimization,'' \emph{arXiv preprint arXiv:1412.6980}, 2014.

\end{thebibliography}

\end{document}